\documentclass[prl,showpacs,twocolumn,amssymb,amsmath]{revtex4}
\usepackage{bm}
\usepackage{graphicx}

\begin{document}
\title{Half-vortices in polariton condensates}

 \author{Yuri~G.~Rubo}
 \email{ygr@cie.unam.mx}
 \affiliation{Centro de Investigaci\'on en Energ\'{\i}a,
 Universidad Nacional Aut\'onoma de M\'exico, Temixco,
 Morelos 62580, Mexico}

\date{May 4, 2007}

\begin{abstract}
It is shown that vortices in linearly polarized polariton
condensates in planar semiconductor microcavities carry two winding
numbers $(k,m)$. These numbers can be either integer or half-integer
simultaneously. Four half-integer vortices $(1/2,1/2)$,
$(-1/2,-1/2)$, $(1/2,-1/2)$, and $(-1/2,1/2)$ are anisotropic,
possess the smallest energy, and define the Kosterlitz-Thouless
transition temperature. The condensate concentration remains finite
within the core of half-vortex and the condensate polarization
becomes fully circular in the core center.
\end{abstract}


\maketitle


\emph{Introduction.}---Recent observations of exciton-polariton
condensation in semiconductor microcavities \cite{Kasprzak,Snoke}
revealed the formation of condensates with a well-defined linear
polarization. The polarization is build-up as a result of
minimization of the energy of polariton-polariton repulsion
$H_\mathrm{int}$ \cite{Laussy06,Shelykh06},
\begin{equation}
 \label{IntEnergy}
 H_\mathrm{int} = \frac{1}{2}\int\!\! d^2r  \{
 (U_0-U_1)(\bm{\psi}^*\!\cdot\bm{\psi})^2
 + U_1\left|\bm{\psi}^*\!\times\bm{\psi}\right|^2      \}.
\end{equation}
Here the integration is over the microcavity plane within the
excitation spot and the polariton condensate wave function (the
order parameter) is written as a complex two-dimensional vector
$\bm{\psi}(\mathbf{r})$. This vector describes the in-plane
component of electric field of the polariton condensate and it is
normalized to the condensate concentration
$n=(\bm{\psi}^*\!\cdot\bm{\psi})$. The polariton repulsion is
characterized by two interaction constants, $U_0$ and $U_1$
\cite{Shelykh06}. They are typically related as $U_0>U_1>0$, so that
at a fixed polariton concentration the minimum of $H_\mathrm{int}$
is reached at $\bm{\psi}^*\!\times\bm{\psi}=0$, i.e., at a linear
polarization.

While the condensation has been observed in \cite{Kasprzak,Snoke}
for the case of localized polaritons, it is of key importance to
understand the structure and polarization properties of topological
defects (vortices) in uniform polariton condensates. First, it is
vortex-antivortex unbinding that defines the critical temperature
$T_c$ for the condensation transition in two-dimensions (2D)
according to the Kosterlitz-Thouless scenario \cite{KostThouless}.
And the estimations of $T_c$ given till now \cite{TCestim} did not
take into account the polarization degree of freedom of polaritons.
Secondly, since vortices are topologically stable objects, they can
be used as long-living optical memory elements. The vortices studied
below can be applied for polarization sensitive optical computing.

\begin{figure}[b]
\includegraphics[width=3.1in]{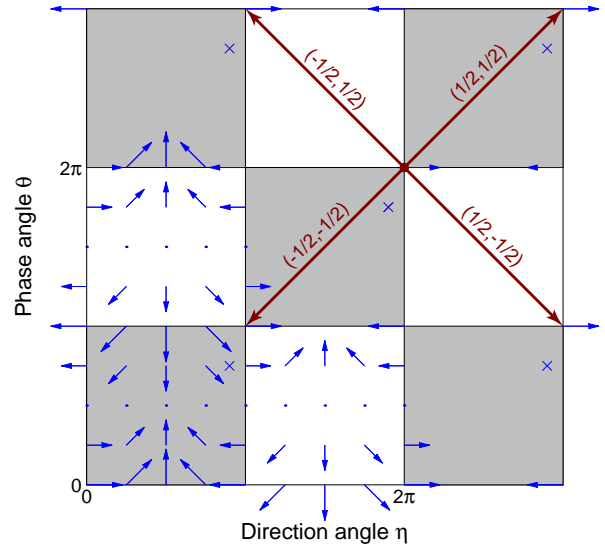}
\caption{\label{OPspace-fig1} Illustrating the order parameter space
of polariton condensate. Blue arrows show
$\mathrm{Re}\{\bm{\psi}\}$. Similar points in the black or white
squares (as, e.g., the points marked with small crosses) represent
identical values of the order parameter. Thick red arrows show four
topologically different changes of the order parameter when one
encircles the cores of four basic half-vortices.}
\end{figure}

\emph{Winding numbers and vortex interactions.}---I begin with the
classification of vortices in polariton condensates. Following the
general approach \cite{Mermin79}, it is necessary to distinguish
topologically different mappings of a closed path in the order
parameter space into a closed path around the vortex core in the
microcavity plane. Since a linear polarization vector can be written
in general as
\begin{equation}
 \label{LinPsi}
 \bm{\psi}_\mathrm{lin}
 =\{\psi_x,\psi_y\}
 =\sqrt{n}\,\mathrm{e}^{\mathrm{i}\theta}\{\cos\eta,\sin\eta\},
\end{equation}
the order parameter is defined by two angles, $\eta(\mathbf{r})$ and
$\theta(\mathbf{r})$. The topology of the order parameter space,
however, does not coincide with the topology of a torus, because the
points $\eta,\theta$ and $\eta+\pi,\theta+\pi$ should be identified.
It is more convenient to imagine the order parameter space as an
infinite chessboard shown in Fig.~\ref{OPspace-fig1}. A vortex is
defined by a line connecting two equivalent points on this
chessboard, and it is characterized by two winding numbers, $k$ and
$m$, such that the angles are changed as $\eta\rightarrow\eta+2\pi
k$, $\theta\rightarrow\theta+2\pi m$ by encircling the vortex core
around. It is seen that the winding numbers $k$ and $m$ can take
either integer or half-integer values together. Basic vortices carry
half-integer values $k,m=\pm 1/2$. Corresponding changes of the
order parameter are shown by four thick arrows in
Fig.~\ref{OPspace-fig1}. Other vortices are reduced to
superpositions of four basic half-vortices.

Existence of half-integer vortices is a general feature of
multi-component condensates \cite{VolovikBook}, but usually they
carry only one half-integer topological charge originating from the
phase angle change. E.g., the half-vortices in $^3$He-A generate
rotations $\mathbf{d}\rightarrow -\mathbf{d}$ of the spin
quantization axis $\mathbf{d}$, accompanied by the phase change
$\theta\rightarrow\theta\pm\pi$ \cite{UsualHV}. Similar
half-vortices have been discussed for the three-component spinor
$s=1$ atomic condensates \cite{Mukerjee}. For a 3D vector
$\mathbf{d}$, different trajectories connecting two antipodes on the
sphere, $\mathbf{d}$ and $-\mathbf{d}$, can be continuously
transformed one to another, so that all half-vortices with the phase
changing by $\pi$ are topologically equivalent. Contrary, the
polariton condensates are two-component, and the clockwise and
counterclockwise rotations of the 2D real vector
$\{\cos\eta,\sin\eta\}$ are topologically distinct. This gives rise
to the second topological charge.

Half-vortices with two winding numbers have being predicted for
$^3$He-A in the parallel-plate geometry \cite{VolovikRMP} where the
spin quantization axis $\mathbf{d}$ is restricted to be
two-dimensional. In this case, however, the sign of phase change is
fixed and only two half-vortices can be present in the same
configuration, while all four half-vortices can coexist in the
polariton condensate.

The total energy of polariton condensate is
\begin{equation}
 \label{TotEnergy}
 H = \int\!\! d^2r                  \left[
 -\frac{\hbar^2}{2m^*}(\bm{\psi}^*\!\cdot\Delta\bm{\psi}) -
 \mu(\bm{\psi}^*\!\cdot\bm{\psi})   \right] + H_\mathrm{int},
\end{equation}
where $m^*$ is the polariton effective mass at the bottom of the
lower polariton branch and $\mu=(U_0-U_1)n$ \cite{Shelykh06} is the
chemical potential. The vortex energy consists of the core energy
and the elastic energy $E_\mathrm{el}$. The latter comes from
polariton kinetic energy far away from the core, where Eq.\
(\ref{LinPsi}) can be used. This gives
\begin{equation}
 \label{ElstEnergy}
 E_\mathrm{el}=\frac{1}{2}\rho_s\int\!\! d^2r
 \left[ (\nabla\eta)^2 + (\nabla\theta)^2 \right]
\end{equation}
with rigidity $\rho_s=\hbar^2n/m^*$. The vortex elastic energies and
vortex interactions in logarithmic approximation can be calculated
from (\ref{ElstEnergy}) similarly to the usual case
\cite{ChaikinLub}.

The energy of a single vortex is
$E_\mathrm{el}^{(\mathrm{s})}=\pi\rho_s(k^2+m^2)\ln(R/a)$, where
$a=\hbar/(2m^*\mu)^{1/2}$ is the core radius (see below) and $R\gg
a$ is the radius of the polariton excitation spot. Clearly, the
elastic energy of the half-vortex is twice smaller than the energy
of a usual vortex [e.g., of the $(0,1)$ vortex]. The energy
$E_\mathrm{el}^{(\mathrm{p})}$ of the pair of vortices $(k_1,m_1)$
and $(k_2,m_2)$ separated by distance $r$, such that $a\ll r\ll R$,
is
\begin{multline}
 \label{PairEnergy}
 E_\mathrm{el}^{(\mathrm{p})} =
 \pi\rho_s[(k_1+k_2)^2+(m_1+m_2)^2]\ln(R/a) \\
 + 2\pi\rho_s(k_1k_2+m_1m_2)\ln(a/r).
\end{multline}
It is seen that only half-vortices with the same sign of the product
$km$ interact. This sign, as we see below, defines the sign of
circular polarization in the center of the vortex core---it is
right-circular for $\mathrm{sgn}(km)=+1$ and left-circular for
$\mathrm{sgn}(km)=-1$. In particular, there is attraction between
the right half-vortex and the right anti-half-vortex, $(1/2,1/2)$
and $(-1/2,-1/2)$, as well as between the left ones, $(1/2,-1/2)$
and $(-1/2,1/2)$. On the same time, the right half-vortices do not
interact with the left half-vortices, and, for example, the energy
of the pair $(1/2,1/2)$ and $(1/2,-1/2)$ simply equals to the energy
of a single $(1,0)$ vortex.\cite{EntropyNote}

At a finite temperature equal numbers of left and right half-vortex
pairs are excited. Since the subsystems of left and right vortices
do not interact, they evolve independently with increasing
temperature and are subject to the Kosterlitz-Thouless transition at
$T_c=(\pi/4)\rho_s$. This temperature is twice smaller than the
critical temperature for spin-less bosons due to the double
reduction of the single half-vortex energy. Clearly, this estimation
of $T_c$ does not take into account the depletion of the condensate,
but it shows the necessity to allow for the polarization degree of
freedom for any realistic calculations of the transition
temperature. The critical temperature is expected to be modified
also by non-parabolicity of polariton kinetic energy, as well as by
the account for the longitudinal-transverse splitting of polariton
branch that lead to the coupling between left and right vortices.
Note also that the symmetry between left and right vortex subsystems
is broken in applied magnetic field.

\emph{Polarization texture of half-vortex core.}---The chemical
potential $\mu$ is found experimentally as a blue-shift of polariton
luminescence line due to the polariton condensation, and, typically,
$\mu\lesssim1\,$meV \cite{Kasprzak,Richard05}. Due to a very small
value of the polariton effective mass $m^*$, ranging from $10^{-4}$
to $10^{-5}$ of the free electron mass \cite{KavokinBook}, the size
of half-vortex core is quite big, $a\gtrsim1\,\mu$m. Therefore, the
polarization texture within the core region of a half-vortex can be
observed by means of near-field luminescence spectroscopy. This
texture is studied in this subsection.

In what follows, only the basic half-vortices with $|k|=|m|=1/2$
will be considered. In this case, the order parameter can be written
in cylindrical coordinates as
\begin{equation}
 \label{HVPsi}
 \bm{\psi}_\mathrm{hv}=\sqrt{n}\,
 \left[\mathbf{A}(\phi)f(r/a)-\mathrm{i}\mathbf{B}(\phi)g(r/a)\right],
\end{equation}
where the azimuthal dependencies are given by
\begin{subequations}
 \label{AandB}
\begin{equation}
 \mathbf{A}(\phi)
 =\mathrm{e}^{\mathrm{i}m\phi}\{\cos(k\phi),\sin(k\phi)\},
\end{equation}
\begin{equation}
 \mathbf{B}(\phi)
 =\mathrm{sgn}(km)\mathrm{e}^{\mathrm{i}m\phi}\{\sin(k\phi),-\cos(k\phi)\},
\end{equation}
\end{subequations}
and the radial functions $f(r/a)$ and $g(r/a)$ are found from
minimization of the vortex energy, i.e., from the equation $\delta
H/\delta\bm{\psi}=0$. Using the relations
\begin{subequations}
 \label{ABprops}
\begin{equation}
 \mathbf{A}^{\prime\prime}(\phi)=\mathrm{i}\mathbf{B}^{\prime\prime}(\phi)
 =-\frac{1}{2}(\mathbf{A}+\mathrm{i}\mathbf{B}),
\end{equation}
\begin{equation}
 \mathbf{A}-\mathrm{i}\mathbf{B}=\{1,\mathrm{sgn}(km)\mathrm{i}\},
\end{equation}
\end{subequations}
one obtains ($\xi=r/a$)
\begin{subequations}
 \label{FGeqs}
\begin{equation}
 f^{\prime\prime}+\frac{1}{\xi}f^\prime-\frac{1}{2\xi^2}(f-g)
 +[1-f^2-\gamma g^2]f=0,
\end{equation}
\begin{equation}
 g^{\prime\prime}+\frac{1}{\xi}g^\prime-\frac{1}{2\xi^2}(g-f)
 +[1-g^2-\gamma f^2]g=0,
\end{equation}
\end{subequations}
where $\gamma=(U_0+U_1)/(U_0-U_1)$.

Equations (\ref{FGeqs}a,b) are symmetric with respect to interchange
of radial functions $f$ and $g$. The half-vortex, however, is
described by an asymmetric solution. Since the polarization should
become linear at large distances, one has $f(\infty)=1$ and
$g(\infty)=0$. This way at $r\gg a$ the solution (\ref{HVPsi})
transforms into (\ref{LinPsi}) with $\eta=k\phi$ and $\theta=m\phi$.
On the other hand, in the half-vortex center one has to demand
$f(0)=g(0)$ in order to remove the divergences produced by terms
$(f-g)/2\xi^2$ in Eqs.\ (\ref{FGeqs}). Taking into account the
relation (\ref{ABprops}b) it is clear that the polarization is fully
circular in the core center ($r=0$), and the sign of circular
polarization is given by $\mathrm{sgn}(km)$, as was mentioned above.

\begin{figure}[t]
\includegraphics[width=3.1in]{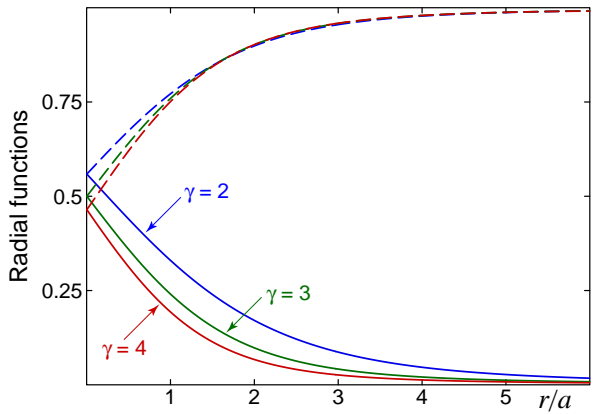}
\caption{\label{FGfig-fig2} Showing the half-vortex radial functions
$f(r/a)$ (dashed) and $g(r/a)$ (solid) for three values of the
interaction parameter $\gamma$.}
\end{figure}

Equations (\ref{FGeqs}) have simple solutions in the particular case
$U_1=U_0/2$, i.e., for $\gamma=3$. Namely, $f(\xi)=[1+h(\xi)]/2$ and
$g(\xi)=[1-h(\xi)]/2$, where $h(\xi)$ is the radial function of a
usual vortex in spin-less condensate \cite{LandavshitsIX}. Since
$h(\xi)$ is monotonously decaying from 1 at $\xi\rightarrow\infty$
to 0 at $\xi=0$, the polariton concentration
$n(r)=f^2(r/a)+g^2(r/2)$ is decreasing from its unperturbed value
$n\equiv n(\infty)$ to $n/2$ in the half-vortex center.

In general, for realistic values of $\gamma$ \cite{gammanote} the
behavior of $f$ and $g$ remains close to the above case. The
numerical solutions of Eqs.\ (\ref{FGeqs}) are shown in Fig.\
\ref{FGfig-fig2}. Note that while the asymptotic behavior of
$f(\xi)$ is $\gamma$-independent for large $\xi$, $f(\xi)\simeq
1-(4\xi^{2})^{-1}$, the second radial function behaves
asymptotically as $g(\xi)\simeq [2(\gamma-1)\xi^2]^{-1}$. Therefore,
the effective size of the half-vortex core grows with decreasing
$\gamma$ and diverges at $\gamma\rightarrow 1$, i.e., when
$U_1\rightarrow 0$. This reflects the fact that polaritons are not
expected to exhibit the superfluid transition for $U_1=0$. In this
case the polariton system has O(4) symmetry and the order is
destroyed at any finite temperature according the nonlinear
$\sigma$-model analysis.

Concerning the polarization texture of the half-vortices given by
Eqs.\ (\ref{HVPsi}) and (\ref{AandB}), it should be noted that
different half-vortices can be transformed to each other by applying
two symmetry operations, (i) the 2D inversion ($x\rightarrow x$,
$y\rightarrow -y$), and (ii) the time inversion or complex
conjugation. Each of these operations used separately transforms the
right $(1/2,1/2)$ half-vortex into one of two left half-vortices,
while the subsequent application of both operations yields the
$(-1/2,-1/2)$ anti-half-vortex. Clearly, these symmetry operations
leave the radial functions $f(r/a)$ and $g(r/a)$ unchanged.

\begin{figure}[t]
\includegraphics[width=2.7in]{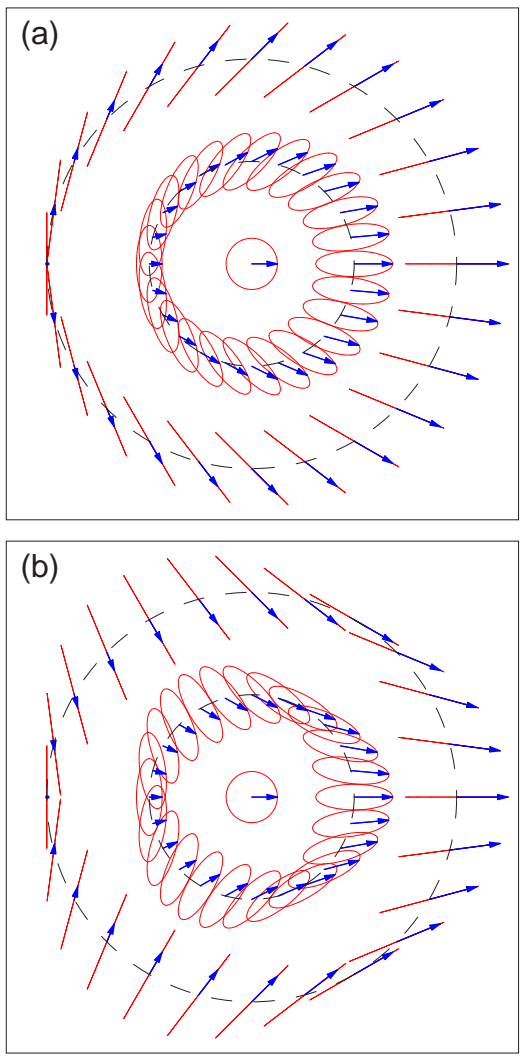}
\caption{\label{HVfig-fig3} Showing two distinct core textures of
basic half-vortices. The case (a) is realized for the right
$(1/2,1/2)$ and the left $(1/2,-1/2)$ half-vortex, while the case
(b)---for right $(-1/2,-1/2)$ and the left $(-1/2,1/2)$ half-vortex.
Arrows indicate the instant polarization at two values of
$r=\mathrm{const}$ (dashed circles) and in the core center $r=0$.
The polarization vectors change in time following red solid lines.}
\end{figure}

The half-vortices possess two anisotropic polarization textures
shown in Fig.\ \ref{HVfig-fig3}(a,b). These figures rely on the
usual representation of the time-dependent electric field as
$\mathbf{E}(t)\propto\mathrm{Re}\{\bm{\psi}\mathrm{e}^{-\mathrm{i}\omega
t}\}$, where $\omega$ is given by the bare frequency $\omega_0$ of
the lower-branch polariton in microcavity blue-shifted by the
chemical potential $\mu$ due to the polariton-polariton repulsion,
$\omega=\omega_0+\mu$. It is seen that when one approaches the core
center linear polarizations convert into elliptical ones. The
circular polarization degree is given by
\begin{equation}
 \label{rhocirc}
 \rho_\mathrm{circ}=\mathrm{sgn}(km)\frac{2fg}{f^2+g^2},
\end{equation}
and $|\rho_\mathrm{circ}|$ increases from 0 to 1 with decreasing
$r$. In spite of the dependence of the directions of main axes of
each elliptical polarization on the azimuthal angle $\phi$, all
elliptical polarizations meet in phase at $r=0$ and transform into
the same circular polarization. Note also that each image in Fig.\
\ref{HVfig-fig3} represents two half-vortices, one with the
clockwise rotation of polarization vector and the other with the
counterclockwise rotation.

The above analysis of half-vortex properties is based on
minimization of the polariton energy subject to specific boundary
conditions, which is valid for the case of polariton condensate in
quasi-thermal equilibrium. Such thermalized polariton condensates
have being recently observed \cite{Kasprzak,Deng2006}. In these
conditions the half-vortex pairs can both appear spontaneously and
be artificially excited. In particular, the half-vortex and
anti-half-vortex pair can be obtained by shining the uniform
linearly polarized condensate with two circularly polarized pulses
having appropriate intensities, spot radii, and separation. Another
option consists in the excitation of a $(0,1)$ vortex using the
optical parametric oscillator setup \cite{Baldassarri}. This vortex
involves only phase change and is topologically equivalent to the
pairs of $(1/2,1/2)$ and $(-1/2,1/2)$ half-vortices, or to the pair
of $(-1/2,1/2)$ and $(1/2,1/2)$ ones, so it can decay in one of
these pairs.

Finally, it should be noted that similar half-vortices with two
topological charges can be present in mixtures of two hyperfine spin
states of $^{87}$Rb atomic condensates \cite{TCatomic}, provided the
order parameter manifold is given by two angles as in Fig.\
\ref{OPspace-fig1}. In this case the half-vortices should also
control the Kosterlitz-Thouless transition observed in these
condensates \cite{KTatomic}.

\emph{Conclusions.}---Polarization vortices in linearly polarized
polariton condensate have been analyzed. It has been shown that four
half-vortices with the direction and phase winding numbers equal to
$\pm 1/2$ have the smallest energy. These half-vortices form two
subsystems exhibiting independent Kosterlitz-Thouless transitions at
the same temperature. The condensate polarization becomes elliptical
within the core of a half-vortex and converts to fully circular in
the core center.

I benefited from discussions with Alexey Kavokin and Michiel
Wouters. This work was supported in part by the grant IN107007 of
DGAPA-UNAM.


\begin{thebibliography}{99}
\bibitem{EntropyNote} In spite of this energy coincidence, the $(1,0)$ vortex
will decay into the pair of half-vortices at any finite temperature
due to the entropy increase. The same applies to the vortices with
higher winding numbers.
\bibitem{gammanote} The interaction parameters $U_0$ and $U_1$ can
be related to the matrix elements of interaction of two polaritons
with the same circular polarization $M_{\uparrow\!\uparrow}$ and
with the opposite circular polarizations $M_{\uparrow\!\downarrow}$.
Namely, $U_0=AM_{\uparrow\!\uparrow}$ and
$U_1=A(M_{\uparrow\!\uparrow}-M_{\uparrow\!\downarrow})/2$, where
$A$ is the normalization area (the excitation spot area). The case
$\gamma=3$ corresponds to the absence of interaction between
polaritons with opposite pseudospins. In reality one expects a weak
attraction between them, so that $\gamma$ is higher than 3.
\end{thebibliography}
\end{document}